# Direct Laser Writing of Surface Micro-Domes by Plasmonic Bubbles

Lihua Dong[1,4], Fulong Wang[1], Buyun Chen[1], Chenliang Xia[1,4], Pengwei Zhu[2], Zhi Tong[2], Huimin Wang[2], LijunYang[3,4], and YuliangWang[1,4,*]

[1] School of Mechanical Engineering and Automation, Beihang University, 37 Xueyuan Rd., Haidian District, Beijing 100191, China

[2] National Center for Materials Service Safety, University of Science and Technology Beijing, Beijing 100083, China

[3] School of Astronautics, Beihang University, Beijing 100191, China

[4] Ningbo Institute of Technology, Beihang University, Ningbo 315832, China

* E-mail: wangyuliang@buaa.edu.cn

**Abstract** Plasmonic microbubbles produced by laser irradiated gold nanoparticles (GNPs) in various liquids have emerged in numerous innovative applications. The nucleation of these bubbles inherently involves rich phenomena. In this paper, we systematically investigate the physicochemical hydrodynamics of plasmonic bubbles upon irradiation of a continuous wave (CW) laser on a GNP decorated sample surface in ferric nitrate solution. Surprisingly, we observe the direct formation of well-defined micro-domes on the sample surface. It reveals that the nucleation of a plasmonic bubble is associated with the solvothermal decomposition of ferric nitrate in the solution. The plasmonic bubble acts as a template for the deposition of iron oxide nanoparticles. It first forms a rim, then a micro-shell, which eventually becomes a solid micro-dome. Experimental results show that the micro-dome radius $R_d$ exhibits an obvious $R_d \propto t^{1/3}$ dependence on time $t$, which can be well interpreted theoretically. Our findings reveal the rich phenomena associated with plasmonic bubble nucleation in a thermally decomposable solution, paving a plasmonic bubble-based approach to fabricate three dimensional microstructures by using an ordinary CW laser.

# 1. Introduction

Due to the enhanced plasmonic effect, noble metal nanoparticles irradiated by a continuous wave (CW) laser can quickly generate a huge amount of heat to vaporize the surrounding liquid, leading to the formation of so called plasmonic bubbles. Plasmonic bubbles have emerged in numerous applications, such as catalytic reaction,[1] biomedical diagnosis and treatment,[2, 3] microfluidics,[4] solar energy harvesting,[5] and solvothermal chemistry.[6] Understanding their nucleation and growth dynamics and exploring the associated phenomena are essential to those applications. The growth dynamics of plasmonic bubbles is strongly influenced by many factors, such as metal nanoparticle arrangement,[7, 8] laser power,[9] gas concentration and pressure,[10, 11] and liquid types.[12] Moreover, it has been revealed that the nucleation of plasmonic bubbles in water can be divided into four phases:[13, 14] a transient and explosively growing bubble (phase 1), an oscillating bubble (phase 2), and two steadily growing life phases (phase 3 and phase 4, dominated by evaporation and gas diffusion, respectively).

The nucleation of plasmonic bubbles is inherently associated with rich physicochemical reactions. Physically, upon laser irradiation on a gold nanoparticle (GNP) decorated sample surface, a strong temperature gradient will be rapidly established in the vicinity of the laser spot area. This hence induces a surface tension gradient, leading to convective flows and Marangoni forces across plasmonic bubbles.[15-18] In this context, a growing plasmonic bubble promotes the directional assembly of micro/nano-materials at solid-liquid interfaces, realizing the formation of micro/nano-structures by taking bubbles as templates. Different materials, like polystyrene particles,[19] quantum dots,[20, 21] Ag/Au particles,[22, 23] conducting polymers,[24] and protein[25, 26] can be accumulated at the three-phase contact lines of plasmonic bubbles by the strong convective flows, leading to the formation of micro/nanostructures. The appearance of these micro/nanostructures can be flexibly controlled by tuning the growth dynamics, [19] size, [23, 24] and locations [27] of the plasmonic bubbles.

Chemically, studies have shown that chemical solvothermal reactions take place along with a growing plasmonic bubble, such as solvothermal decomposition[28, 29] and solvothermal synthesis.[30-35] These reactions are mainly triggered by photothermally induced high temperature around plasmonic bubbles. When the temperature exceeds a threshold value, it will promote the thermal driven reactions of the surrounding liquid, leading to the formation of micro structures along the three-phase contact line at solid-liquid interfaces.

In this paper, instead of microstructure formation at the three-phase contact line, we report the formation of a three-dimensional (3D) micro-dome by taking a plasmonic bubble as the template. Upon laser irradiation of a GNP decorated sample surface in ferric nitrate solution, we observed rich growth dynamics of plasmonic bubbles accompanied with thermal decomposition. Interestingly, the nanoparticles induced by the solvothermal decomposition are attached to the nucleated plasmonic bubbles, leading to the formation of a 3D micro-dome. The detailed process, the mechanism, as well as the controllability of micro-dome formation are systematically investigated by adjusting key control parameters of laser power, ferric nitrate concentration, as well as the magnification of the focusing objective lenses. We envision that our finding highlights the rich physicochemical processes along with plasmonic bubble nucleation and paves a plasmonic bubble assisted approach of 3D microstructure fabrication.

## 2. Results and Discussion

**2.1 Observation of a jet plume and micro-dome formation**

In this study, experiments were conducted in ferric nitrate solution. Compared with the growth dynamics of plasmonic bubbles in water,[13] we observed two new phenomena in the solution, which depend on laser power $P_l$ and ferric nitrate concentration $c$ in the solution. At a relatively higher laser power and ferric nitrate concentration, we observed the first phenomenon, a rapidly rising jet plume along with laser irradiation. **Figure 1a(I)** shows detailed generation process of an exemplary jet plume at $P_l = 48.47$ mW, $c = 10.0\%$, and a focusing objective lens with magnification $m = 20\times$. After a 163 ms delay upon laser irradiation, we observed a small bump like structure at the laser spot ($t = 250$ ms). At about $t = 350$ ms, the bump was ejected from the substrate with a well-defined jet plume structure under the ejected bump. At the top of the jet plume, it exists a shell-like microstructure. Once the laser irradiation was switched off, the plume stopped to grow right away and then gradually faded away. We also find that the height of the jet plume increases with the period of time $t_d$ of laser irradiation. As shown in Figure 1b, the height of the jet plume linearly increases with $t_d$. The details of growth dynamics of the jet plumes can be seen in Movie S1 in the Supporting Information, where five different jet plumes with laser irradiation periods $t_d$ of 0.4, 0.5, 0.6, 0.7, and 0.8 s were presented from side view. In Movie S2, we present another plume from bottom view. From the video, one can clearly see that a micro-shell structure was first ejected from the substrate. After laser irradiation was switched off, the micro-shell structure was left on the substrate.

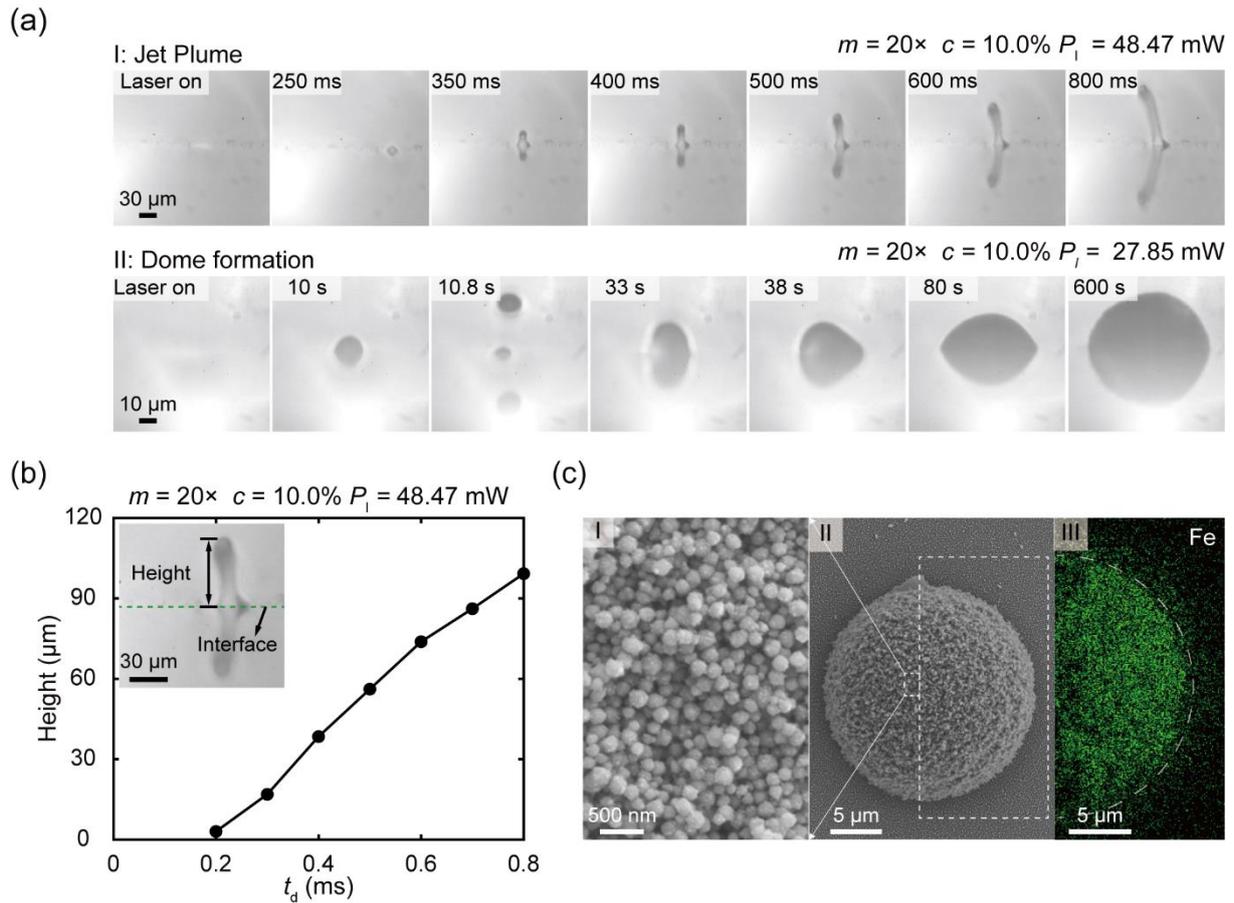

**Figure 1**. Jet plume and dome formation on a GNP decorated sample surface irradiated by a CW laser in ferric nitrate solution. (a) Sequentially captured side view images of a rising jet plume (I) and a micro-dome (II) triggered by the laser irradiation. (b) The height of a jet plume as a function of laser irradiation period $t_d$ for the jet plume shown in figure (a-I). (c, middle) A scanning electron microscope (SEM) image of an obtained micro-dome. The micro-dome exhibits a well-defined hemi-spherical shape. (c, left) The enlarged SEM image for an area selected by a rectangle box in the middle SEM image. It clearly shows that the micro-dome is densely covered with nanoparticles. (c, right) An energy dispersive spectroscopy (EDS) image indicating the distribution of Fe element in a selected area by a dashed box in the right side of SEM image, indicating the particles are actually iron oxide nanoparticles.

After a moderate laser power $P_l$ and ferric nitrate concentration $c$ were applied, we observed the second phenomenon. No ejected jet plume was observed. Instead, we observed the formation of a micro-dome in the laser spot area. The growth process for a micro-dome is depicted in Figure 1a(II) at $P_l$ = 27.85 mW, $c$ = 1.0%, and $m$ = 20×. For details, readers are referred to a synchronized top and bottom view video (Movie S3) and Figure S1 in the Supporting Information. After a delay from the beginning of the laser irradiation, a wiggling bubble first appeared. The synchronized top and bottom view video shows that the bubble is covered by black particles. The nucleated bubbles detached from the substrate and a new wiggling bubble repeatedly appeared. This process may repeat for several times. Gradually, a bubble was stabilized without wiggling motion on the substrate. Initially, the stabilized bubble laterally expended to a size much larger than the initially appeared wiggling bubble. It was semi-transparent and one can see the quickly moving particles over the bubble surface (bottom view video around 56.0 s in Movie S3). With time, the bubble became opaque and a micro-dome was formed. The size of the micro-dome gradually increased with a slightly increasing contact angle. Once the laser was switched off ($t$ = 600 s), some particles slipped off from the micro-dome. After that, the micro-dome remained unchanged on the substrate.

The middle figure in Figure 1c depicts a scanning electron microscope (SEM) image of an obtained micro-dome. It clearly shows that a well-defined micro-dome is obtained on the substrate. The left figure in Figure 1c shows an enlarged SEM image of a selected area by a rectangular box in the middle figure. Interestingly, it shows that the micro-dome is densely covered with nanoparticles, which are supposed to be iron oxide particles obtained by solvothermal decomposition of ferric nitrate in the solution. Ferric nitrate can be decomposed into $NO_2$, $O_2$ and iron oxide particles when temperature is over 348 K in water.[36] To further verify that the nanoparticles are actually iron oxide, we ran an energy dispersive spectrometer (EDS) scanning. The right figure in Figure 1c shows the distribution of Fe element in an area selected by a dashed box in the middle figure. It clearly shows that Fe is densely distributed over the micro-dome surface, indicating that the micro-dome is covered by iron oxide nanoparticles. If laser power $P_l$ and ferric nitrate concentration $c$ were further decreased, no apparent reaction was observed.

## 2.2 Mechanism of micro-dome formation

The laser irradiation period $t_d$ was adjusted to reveal the underlying mechanism of micro-dome formation. **Figure 2a(I)** shows the top view optical images of the finally obtained microstructures at solid-liquid interfaces with different $t_d$. The results show that the shape of the obtained microstructures changes from a micro-rim ($t_d$ =50 s), to semi-closed microstructures ($t_d$ = 150, 180, and 200 s), and eventually to completely enclosed micro-domes ($t_d$ = 400 and 600 s) with increased $t_d$. Figure 2a(II) shows corresponding isometric view SEM images of the same microstructures in Figure 2a(I). All microstructures were produced with a 20× objective lens at $P_l$ =27.85 mW and $c$ = 1.0%. For synchronized side view and bottom view optical images, as well as the isometric and top view SEM images of the microstructures, readers are referred to Figure S2 for details.

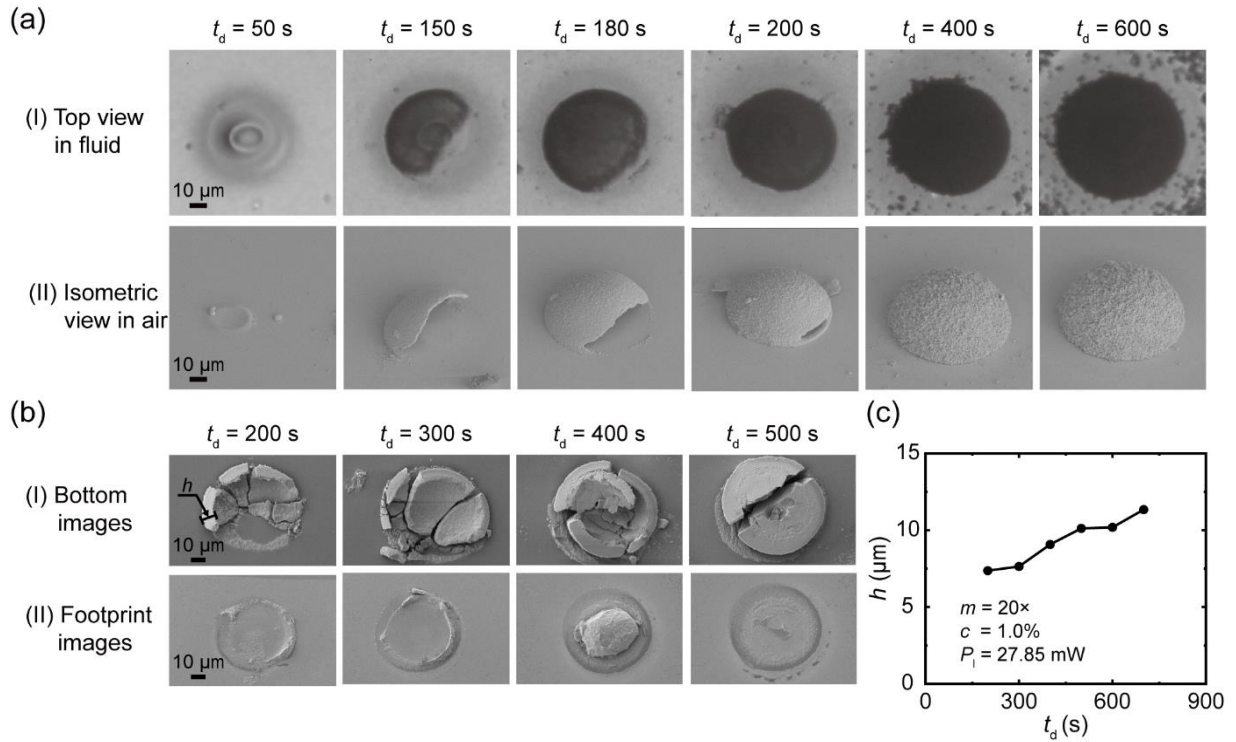

**Figure 2**. SEM and optical images revealing the step by step formation of micro-domes. (a) Top view optical images (I) and isometric view SEM images (II) of several microstructures obtained with different laser irradiation period $t_d$. The microstructures started from a rim ($t_d$: 50 s), then to semi-closed micro-shells ($t_d$: 150 s, 180 s, and 200 s), and eventually enclosed micro-domes ($t_d$: 400 s and 600 s). (b) SEM images of flipped over micro-structures (I) and corresponding SEM images of the footprint areas after micro-domes were removed from the substrate (II). From $t_d$ = 400 s, completely solid domes are obtained.

(c) The thickness of the micro-shells $h$ as a function of $t_d$. Even micro-structures turned into solid micro-domes, one can still tell the existence of micro-shells, of which thickness increases with $t_d$.

For a short laser irradiation period of $t_d$ = 50 s, after intermittently detached wiggling bubbles, we obtained a gradually stabilized bubble with a semi-transparent interface. Beneath the bubble, we observed a micro-ring structure on the substrate ($t$ = 50 s in Movie S3). Once laser irradiation was switched off, the semi-transparent interface disappeared right away with scattering particles around the laser spot. Only a micro-rim structure remained on the substrate (Figure 2a(I), $t_d$ = 50 s). The micro-rim structure can also be well recognized in the SEM image in Figure 2a(II) ($t_d$ = 50 s). We systematically investigated the formation of the micro-rims and found that they are confined by laser spot size. During laser irradiation, rims gradually increased with time with their maximum size equals to that of the laser spot on the substrate. For details, readers are referred to Figure S3.

Next, $t_d$ was increased to 150 s. At beginning, it follows exactly the same phenomenon to the case of $t_d$ = 50 s. Once laser irradiation was switched off, the microstructure was not completely disappeared (Figure S4 and Movie S4). After particle scatting, about half of dome structure remained on the substrate (Figure 2a(I), $t_d$ = 150 s). The isometric view SEM image (Figure 2a(II), $t_d$ = 150 s) clearly shows that a shell structure remained on the substrate. This indicates that there was an enclosed shell before the laser irradiation was switched off. Once laser was switched off, part of micro-shell disappeared with scattering particles. When $t_d$ was increased to 180 s, the major portion of the dome remained on the substrate after the laser irradiation was switched off. For $t_d$ = 200 s, the dome was almost enclosed. Only a very tiny part remained open. For further increased $t_d$ of 400 and 600 s, the obtained micro-domes are completely enclosed. Moreover, their lateral size increases with $t_d$.

We also investigated how the inner part of the micro-domes evolve with $t_d$. By flipping over the obtained micro-domes, the inner part of these micro-domes as well as their footprint areas on the substrate can be characterized, as shown in Figure 2b(I) and (II). From the SEM images, one can see that the micro-domes are shell structures when $t_d$ is lower than 300 s. With increasing $t_d$, the micro-domes became solid micro-structures and their inner parts were filled with particles. We measured the thickness of micro-shell $h$ for micro-domes obtained with different $t_d$. The results are shown in Figure 2c. It shows that $h$ increases with $t_d$. From the flipped over micro-domes, one can see that initially a stabilized bubble acts as the template to form a shell structure. With time, the hollow part of the micro-shell is filled with iron oxide particles and the whole micro-dome becomes a solid structure.

From the above analysis, we propose the following mechanism of micro-dome formation, as illustrated in **Figure 3**. Basically, it involves three major phases: micro-rim (I-III), micro-shell (IV-VI), and micro-dome (VII-IX). In the first phase, the liquid temperature in the vicinity of the laser spot is rapidly elevated upon laser irradiation. This leads to the solvothermal decomposition of ferric nitrate solution, which produces the mixture of the $NO_2$, $O_2$ and iron oxide nanoparticles. The produced gas facilitates the formation of a wiggling bubble, of which lateral size is confined by the laser spot size. Because the three-phase contact line exhibits the highest temperature across the bubble surface, the solvothermal decomposition mostly takes place at the three-phase contact line. That is why we observed the formation of a rim structure (Figure 3(I)). This process is similar to that reported somewhere else.[28, 34]

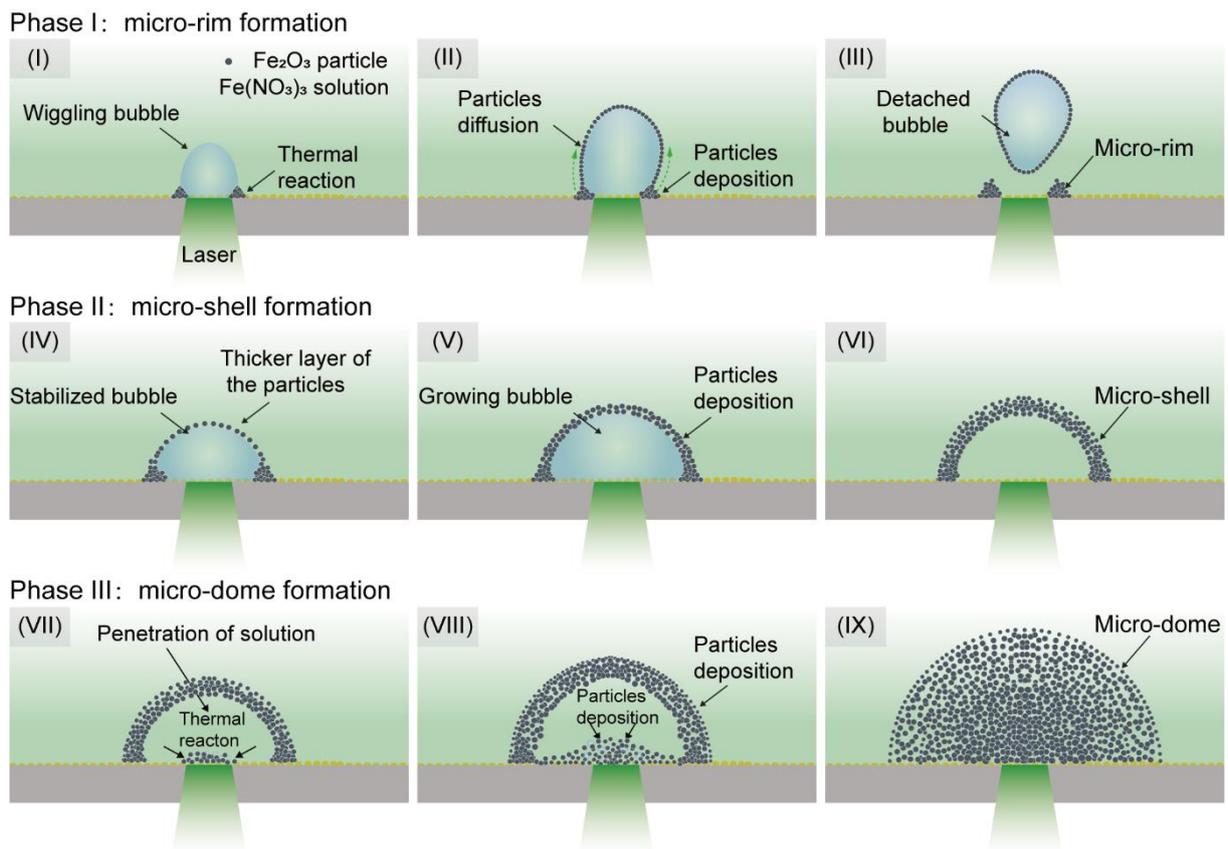

**Figure 3**. Schematic diagrams indicating the detailed process of micro-dome formation by taking a plasmonic bubble as a template. It includes three major phases. The first phase is the formation of a micro-rim along the three-phase contact line of intermittently generated wiggling bubbles (I - III). The second phase is the formation of a micro-shell by taking a stabilized plasmonic bubble as the template (IV - VI). The third phase is steadily growth inside and outside the micro-shell structure to eventually form a solid micro-dome structure (VII - IX).

Moreover, because the temperature at the substrate is much higher than that at the apex of the bubble (Figure S5a), there exists an upward convective flow along the surface of the bubble, which brings iron oxide nanoparticles to the upper part of the bubble surface (Figure 3(II)). For details, readers are referred to Figure S5(b) for a thermal gradient induced convective flow obtained by a numerical simulation in COMSOL software. Once the nanoparticles contact the bubble interface, they will be captured and deposited on the bubble surface. The unstable wiggling bubble can easily detach from the substrate (Figure 3(III)). The nucleation and detachment of wiggling bubbles can repeat for several times. During the process, the size of the rim gradually increases.

In the second phase, the wiggling bubble is gradually stabilized at the interface (Figure 3(IV)). The solvothermal decomposition continues due to the sustained laser energy input. Nanoparticles are attached to the stabilized bubble, forming a thin nanoparticle shell across the bubble surface (Figure 3(V)). At this moment, nanoparticles are not firmly bonded to each other. If laser irradiation is switched off, the bubble will fall apart with scattered nanoparticles around the laser spot. With time, more and more nanoparticles are attached to the bubble surface, leading to a thicker nanoparticle shell (Figure 3(VI)). Due to the higher inner pressure and high local temperature, nanoparticles in the shell gradually fused into a solid shell. As a result, even laser irradiation is switched off, the shell structure does not fall apart. Only a portion of newly attached/nucleated nanoparticles fall off from the shell structure.

In the third phase, the gas-liquid interface will disappear with an established solid shell. The stabilized bubble completely turns into a solid shell structure (Figure 3(VII)). Ferric nitrate solution permeates the micro-shell and enters the internal hollow part. Through a temperature simulation across a solid shell structure (see Figure S5c for details), one can see that the temperature of the micro-shell surface is about 360 K, which exceeds the required temperature for ferric nitrate decomposition. As a result, new nanoparticles keep forming at the outer surface and internal hollow part. Both the internal and external parts of the micro-shell grow simultaneously (Figure 3(VIII)). Additionally, once the liquid-gas interface disappears, convective flow will be significantly decreased (Figure S5d). Therefore, nanoparticle motion induced by the convective flow does not play any role in this phase. Instead, the direct nucleation of nanoparticles on the micro-shell surface governs the growth dynamics of micro-domes. With time, more and more nanoparticles are generated inside the micro-shell, leading to the formation of a completely solid micro-dome (Figure 3(IX)).

The above revealed mechanism of micro-dome formation can also explain the formation of jet plumes mentioned earlier. A higher laser power leads to an accelerated thermal decomposition. As a result, a solid-shell is quickly formed. Inside the micro-shell, the violent solvothermal decomposition takes place and a huge amount of gas is quickly generated, leading to a rapidly increasing pressure. Once the pressure is above a threshold value, the formed micro-shell will be ejected, forming a jet plume.

**2.3 Modeling of growth dynamics of micro-domes**

Once a stable micro-shell is formed, the solvothermal decomposition of ferric nitrate solution continues to take place across the micro-shell surface. As a result, newly nucleated iron oxide nanoparticles are attached on the micro-shell surface, leading to the increased size. Here we investigate the growth dynamics of micro-domes in this process. A 50× objective lens was applied for this purpose. Compared with the 20× objective lens, the 50× objective lens provides a smaller laser spot size and hence higher laser power density. As a result, the processes of wiggling bubble and micro-shell formation in the early two phases are significantly shortened. This leads to a highly repeatable growth dynamics of micro-domes, as shown in Figure S6.

**Figure 4a** shows the radius of curvature $R_d$ of repeatedly produced micro-domes as a function of time $t$ along with laser irradiation at three different concentration $c$ of 0.3%, 0.5%, and 1.0% for a fixed laser power of 7.85 mW by using the 50× focusing objective lens. It clearly shows that when $t$ is less than 200 s, $R_d$ exhibits a clear dependence on $t$ as $R_d \propto t^{1/3}$. This dependence is further verified on the final size of the obtained micro-domes with different laser irradiation period $t_d$, as shown in Figure 4b.

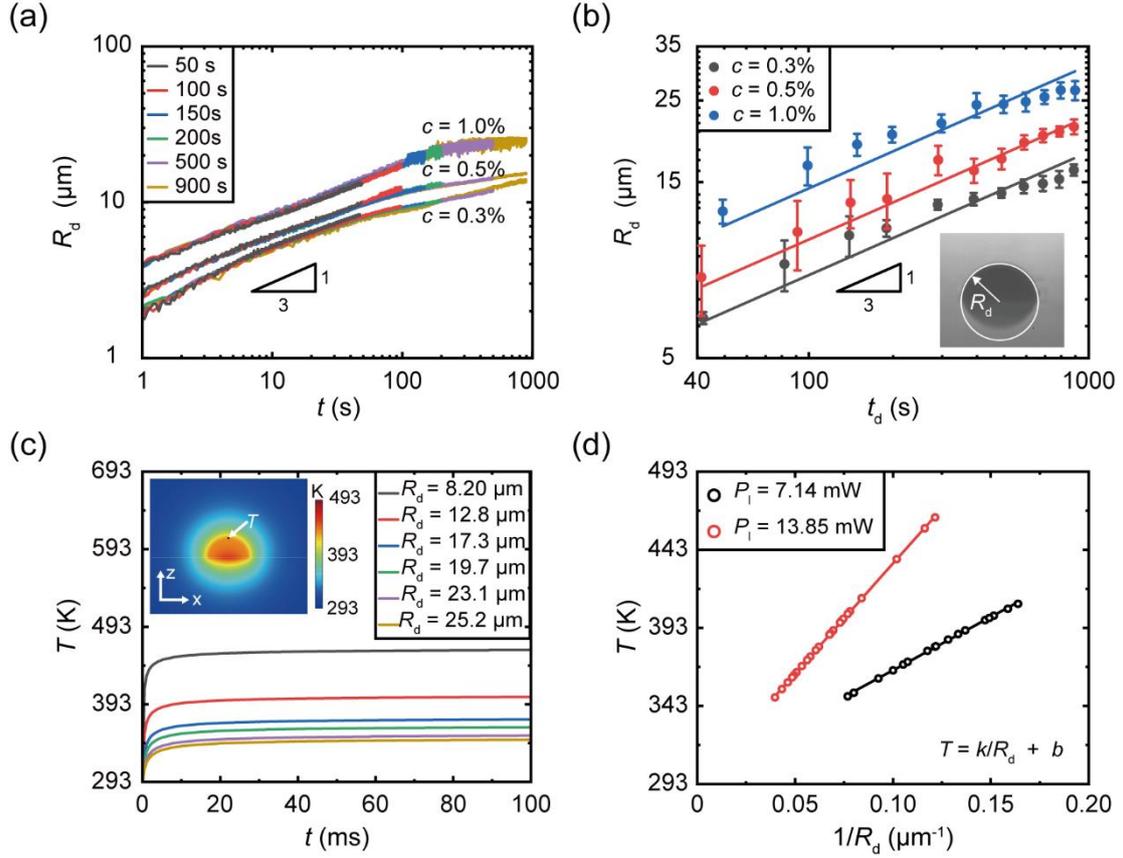

**Figure 4**. Growth dynamics of micro-domes obtained with a 50× objective lens. (a) Radius of curvature $R_d$ of micro-domes as a function of $t$ along laser irradiation in a double logarithmic plot. It exhibits a clear $R_d \propto t^{1/3}$ dependence for $t \leq 200$ s. (b) The final size of the micro-domes obtained with different laser irradiation periods $t_d$ after laser irradiation was switched off. The same $R_d \propto t_d^{1/3}$ dependence is obtained. (c) The temperature as a function of time at apex of micro-domes obtained from a numerical simulation of temperature field due to sustained laser input for micro-domes with different radius of curvature $R_d$. The inset is a constructed temperature field. The temperature rapidly reaches a steady value in less than 10 ms upon laser irradiation. (d) The numerically obtained steady temperature $T$ at the apex of micro-domes as a function of $1/R_d$. $T$ roughly linearly increases with $1/R_d$ under the same laser power $P_l$.

The observed $R_d \propto t^{1/3}$ dependence is quite interesting. This is a typical production-limited growth under a constant growth rate for a spherical object,[14] namely $dV_d/dt = b$, where $V_d$ is the volume of micro-domes and $b$ is a pre-factor. This leads to the following relationship:

$$V_d(t) \propto t \text{ or } R_d(t) \propto t^{1/3} \qquad (1)$$

In this case, the growth of micro-domes is because of the solvothermal decomposition of ferric nitrate in the solution. This requires sustained energy input from laser irradiation. The constant production rate $dV_d/dt$ indicates that the heat conversion from laser energy to the thermal energy at the surface of micro-domes remains constant during the process. Here, this is theoretically and numerically confirmed.

Theoretically, we first assume a steady temperature field during the micro-dome formation, i.e., $\partial_t T = 0$. This leads to the following linear Fourier equation of heat conduction:

$$\partial_t T(r,t) = \kappa \frac{1}{r^2} \partial_r (r^2 \partial_r T(r,t)) = 0 \tag{2}$$

where $\partial_t T(r, t)$ is the temperature gradient in a micro-dome along its radial direction and $r$ is the radial coordinate from the center of the micro-dome, and $\kappa$ is the thermal diffusivity of iron oxide. By solving Equation 2 at $r = R_d$, we obtained,

$$\partial_r T \big|_{r=R_d} = \frac{-k}{R_d^2} \tag{3}$$

where $\partial_r T|_{r=R_d}$ is the temperature gradient of the micro-dome at the micro-dome surface and $k$ is a pre-factor.

This $\partial_r T|_{r=R_d} \propto R_d^{-2}$ dependence was further confirmed through a numerical simulation of temperature evolution around a solid micro-dome by using the COMSOL software. In the simulation, we applied the same parameters as used in experiments. One constructed temperature filed is shown in the inset figure of Figure 4c. With the constructed temperature fields, we measured the temperature at apexes of micro-domes. The measured temperature $T$ as a function of laser irradiation time $t$ for micro-domes with different $R_d$ at $P_l$ = 13.85 mW is shown in Figure 4c. The results show that $T$ rapidly increases with $t$ upon laser irradiation and reaches a steady value at about $t$ = 10 ms. The required time to reach steady temperature field is at least three orders of magnitude shorter than that for micro-dome formation. This indicates that the assumption of steady state temperature field used in Equation 2 is valid. Moreover, numerical simulation shows that the steady value of $T$ at micro-dome surface decreases with increasing $R_d$. We further investigate how $T$ changes with $R_d$. The results are shown in Figure 4d. One can see that the steady temperature $T$ linearly increases with $1/R_d$, namely $T \propto 1/R_d$. Consequently, we have $\partial_r T \propto 1/R_d^2$, which is consistent with Equation 3. Additionally, results show that a higher $P_l$ leads to a higher steady temperature $T$ for a given $R_d$.

With the obtained temperature gradient Equation 3 at the micro-dome interface, one can obtain the heat flow rate $q$ over micro-dome surface, which drives the solvothermal decomposition at the micro-dome surface.

$$q = \kappa \frac{\partial T}{\partial r}\bigg|_{r=R_d} 2\pi R_d^2 \tag{4}$$

The following chemical reaction (Equation 5) takes place on micro-dome surface when $T$ exceeds a threshold value, resulting in the formation of iron oxide nanoparticles.

$$4Fe(NO_3)_3 \xrightarrow{\Delta} 2Fe_2O_3 + 12NO_2 \uparrow + 3O_2 \uparrow \tag{5}$$

Here we assume the conversion efficiency $f$ (with $0 < f < 1$) from $q$ to solvothermal decomposition remains constant and the assumed heat of solvothermal decomposition for 1 mole ferric nitrate equals to $\Delta H$. From the Hess's law, the production rate of iron oxide particles in unit time can be given as,

$$\dot{m}_d = \frac{fq}{2\Delta H} M \tag{6}$$

where $m_d$ is the mass of the micro-dome and $M$ is the relative molecular weight of iron oxide.

The mass growth rate of the micro-dome can also be expressed as the following equation by taking it as half sphere,

$$\dot{m}_d = \rho \dot{V}_d = 2\pi \rho R_d^2 \dot{R}_d \tag{7}$$

where $V_d$ is the volume of the micro-dome and $\rho$ is the density of iron oxide. By combining Equations 3, 4, 6, and 7, we can easily obtain the following relationship:

$$dR_d^3 = \frac{3\kappa kfM}{\Delta H \rho} dt \tag{8}$$

From Equation 8, one can easily obtain the relationships shown in Equation 1, which is consistent with our experimental results shown in Figure 4(a), 4(b), and Figure S6(b).

We also investigated how iron oxide nanoparticles on micro-dome surfaces evolves with time. We choose the same ferric nitrate concentration $c = 0.3\%$ and laser power $P_l = 4.05$ mW. **Figure 5a(I)** and 5a(II) present SEM images of iron oxide nanoparticles captured by the 50× and 100× objective lenses, respectively. Figure 5b shows how the average size of nanoparticles changes with $t_d$. It clearly shows that nanoparticle radius $R_p$ increases with $t_d$ for each objective. This is believed to be due to the Ostwald ripening. As the solvothermal decomposition prolonging, these nanoparticles would experience an Ostwald ripening process,[37, 38] leading to the increased size of nanoparticles. The pre-existed small nanoparticles act as nuclei for newly produced iron oxide particles due to their high surface energy. Simultaneously, the small particles have the fast dissolving rate. Under the dissolution - precipitation mechanism, the ferric oxide nanoparticles grow bigger as the thermal reaction goes forward.

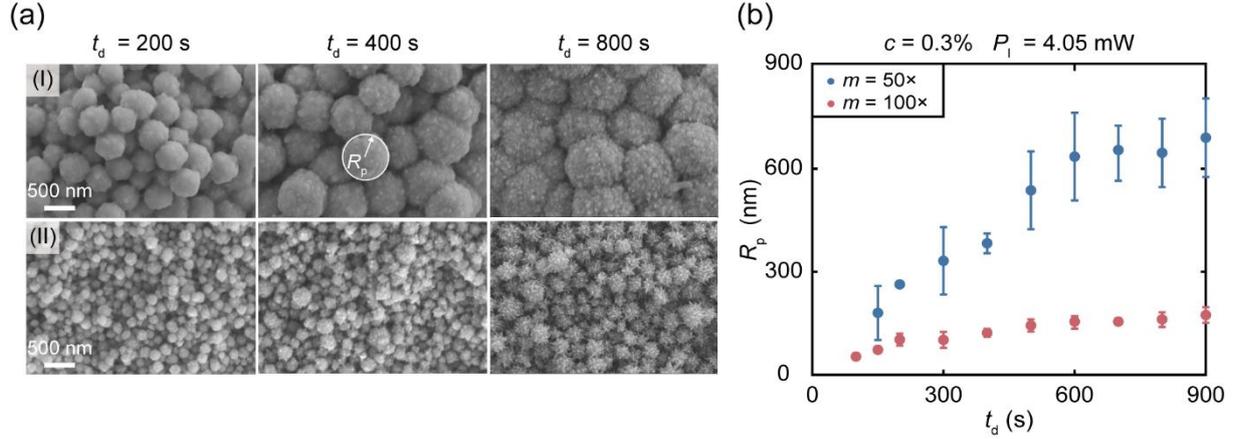

**Figure 5**. Evolution of iron oxide nanoparticles along with growing micro-domes. (a) SEM images of iron oxide nanoparticles on micro-dome surfaces by using a 50× (I) and 100× (II) objective lenses at $P_l$ = 4.05 mW and $c$ = 0.3%. The size of nanoparticles increases with $t_d$. (b) Radius $R_p$ of nanoparticles as a function of laser radiation period $t_d$ obtained by the 50× and 100× objectives. $R_p$ increases with $t_d$.

Moreover, the nanoparticle obtained by the 50× objective lens is much larger than that of the 100× objective lens. This may be because the solvothermal decomposition is much more violent with the 100× focusing lens than that of the 50× lens. This results in the rapid nucleation and hence a much higher concentration of iron oxide nanoparticles. As a result, the particle nucleation rate is higher compared to the Ostwald ripening dominant growth of the nanoparticles with the 50× objective lens. The nucleation of new nanoparticles dominates the process.

**2.4 Controllability of the produced micro-domes**

In this part, we first explore the available range of control parameters for the formation of micro-domes. After that, the controllability of micro-domes is investigated by tuning different control parameters. As mentioned earlier, there are three typical experimental phenomena. Here we quantitatively investigated how the phenomena depend on the selected control parameters, namely laser power $P_l$, ferric nitrate concentration $c$, and magnification $m$ of the focusing objectives. **Figure 6a** shows the result for the 20× objective lens. The parameter space can be divided into three regimes: no reaction, dome formation, and jet plume formation. For a specific concentration $c$, there is no reaction for a low laser power $P_l$. This is because the induced temperature is not high enough to trigger thermal decomposition. With increasing $P_l$, it first enters the regime of dome formation. The threshold value to trigger micro-dome formation decreases with increasing $c$. This should be because the increased concentration of ferric nitrate facilitates thermal decomposition, leading to decreased threshold temperature and laser power $P_l$. If we further increase $P_l$, it enters the regime of jet plumes. The high laser power leads to accelerated thermal decomposition. As a result, a micro-shell is quickly formed. Inside the micro-dome, gas is violently produced. With increasing inner pressure, the micro-shell is ejected from the substrate, leading to the formation of a jet plume. We also find that the threshold value of $P_l$ to trigger jet plume formation remains almost constant with $c$.

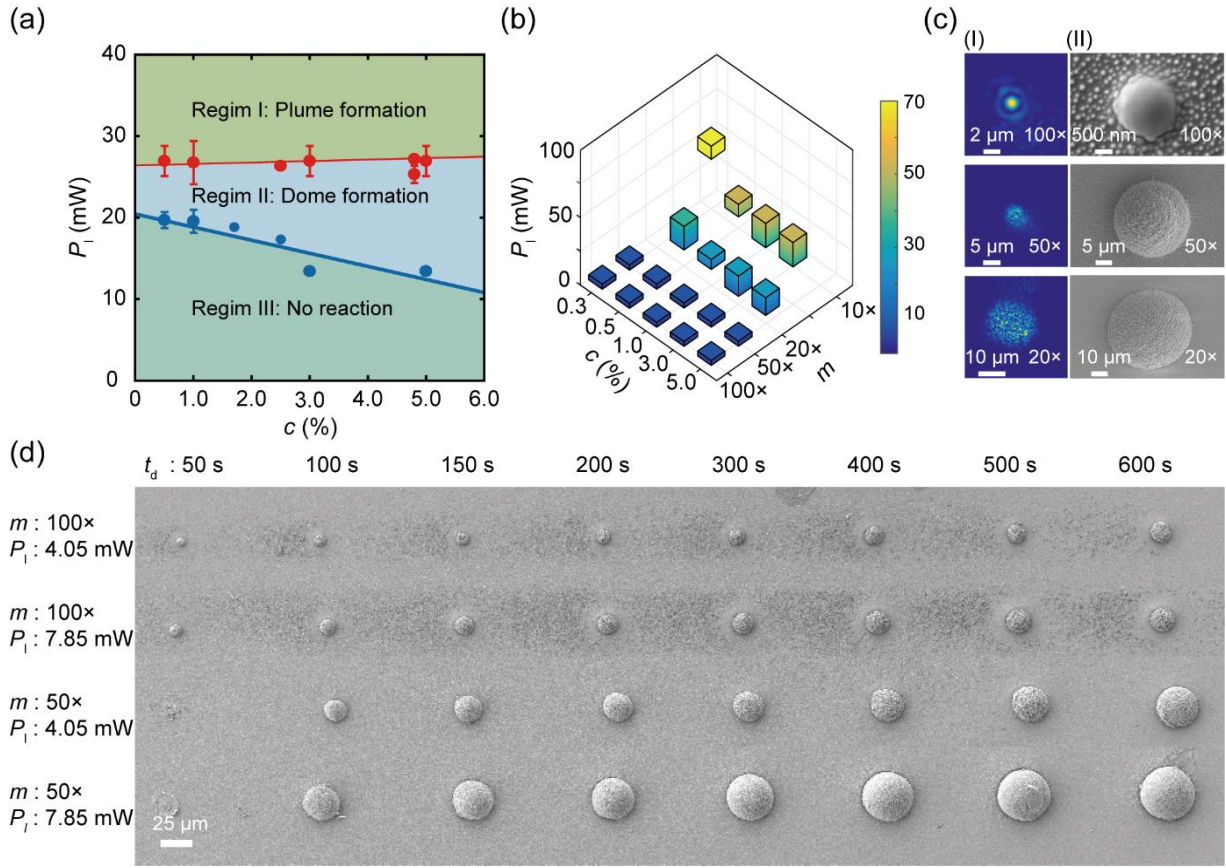

**Figure 6**. The dependence of micro-dome formation on the control parameters. (a) The $c$ - $P_l$ parameter space with the 20× objective lens can be divided into three regimes: no reaction, dome formation, and jet plume formation. (b) Determined regimes of micro-dome formation in the $c$, m, and $P_l$ 3D parameter space. Each bar represents the available range of $P_l$ to trigger micro-dome formation. (c, I) Optical images of laser spot areas and (c, II) micro-domes fabricated with the 100×, 50×, and 20× objective

lenses. The diameters of the laser spots are 5.0, 11.0, and 40.0 $\mu m$ for the 100×, 50×, and 20× objective lenses, respectively. (d) An array of fabricated micro-domes with increasing size produced by the 50× and 100× objective lenses at the same ferric nitrate concentration of 1.0%.

Besides the laser power, the magnification $m$ of the focusing objective lenses can significantly influence the temperature field in the vicinity of laser spots.[39] As a result, the division of the three regimes is strongly related to $m$. Figure 6b shows the results of the determined regimes of dome formation for objective lenses with different values of $m$. In the figure, each bar represents the corresponding range of $P_l$ to trigger micro-dome formation at a specific combination of $c$ and $m$. It clearly shows that the range of $P_l$ is enlarged with decreasing $c$ and decreasing $m$. Note that, for a low value of $c = 0.3\%$, there is no reaction for the 10× and 20× objective lenses.

Regarding the controllability of micro-dome size, we have shown that they can be well tuned by adjusting $t_d$ (Figure 4(a, b)). Since initially microbubbles act as templates for micro-dome formation, one can expect that the size of micro-domes can be well controlled by adjusting the size of microbubbles, which is directly related to the laser spot area. Figure 6c(I) presents the optical images of laser spot areas obtained with different focusing objective lenses. The diameters of the laser spot areas are 5.0, 11.0, and 40 $\mu m$ for 100×, 50×, and 20× objective lenses, respectively. The objective lenses with different $m$ can produce microbubbles with a large size distribution. Consequently, we can tune the size of micro-domes in a large range. As shown in Figure 6c(II), a micro-dome with a 1.7 $\mu m$ diameter was obtained with the 100× objective lens. For an objective lens with a low value of $m = 10\times$, large micro-domes with a diameter over 100 $\mu m$ can be readily obtained. Besides, the size of microbubbles can also be well tuned by adjusting laser power $P_l$.[14] In Figure 6d, we present an obtained micro-dome array by adjusting $m$, $t_d$, and $P_l$. In the array, the diameters of micro-domes were adjusted from 5.0 $\mu m$ to 40.0 $\mu m$. This demonstrates a good controllability of micro-domes.

## 3. Conclusion

To summarize, we have shown that there are three typical physicochemical responses of no reaction, dome formation, and jet plume formation on a laser irradiated and GNP decorated substrate in ferric nitrate solution. The detailed process of micro-dome formation was revealed.

It goes through three major phases, namely, micro-rim, micro-shell, and solid micro-dome. Upon laser irradiation, solvothermal decomposition takes place, leading to the formation of wiggling microbubbles and nucleation of iron oxide nanoparticles. The nanoparticles first aggregate at the three-phase contact lines of the repetitively detached wiggling microbubbles, resulting in the formation of a rim structure. With time, a gradually stabilized wiggling bubble acts as a template for the deposition of nanoparticles, forming a hollow micro-shell. Solvothermal decomposition continues to take place at the outer surface and inner part of the hollow micro-shell and a solid micro-dome is then formed. Both experimental results and a developed theoretical model show that the radii $R_d$ of micro-domes exhibits an obvious $R_d \propto t^{1/3}$ dependence on time $t$, indicating a constant production rate of iron oxide nanoparticles across micro-dome outer surface under a given laser power $P_l$. By adjusting the key control parameters, the diameters of micro-domes can be well tuned from 1 to 100 $\mu$m, demonstrating a good controllability.

## 4. Experimental Section

*Optical setup*

**Figure 7a** shows a schematic diagram of the optical setup developed for experiments. In the setup, the gold nanoparticle (GNP) decorated sample surface was placed in a cuvette. The cuvette was completely filled with ferric nitrate (Fe(NO$_3$)$_3$) solution, which was prepared by dissolving ferric nitrate nonahydrate particles (Fe(NO$_3$)$_3$·9H$_2$O, Sigma Aldrich) in deionized water (18.2MΩ). A continuous wave (CW) laser with a wavelength of 532 nm was used for irradiation of the sample surface. The laser power on the sample surface was tuned *via* a polarizer and a half-wave plate. The laser power was measured by a photodiode power meter. Two high-speed cameras (Nova & SAZ, Photron, Japan) were used to implement synchronized imaging of laser spot areas for bubble nucleation and micro-dome formation from bottom view and side view. A 50× long working distance objective lens (LMPLFLN, Olympus) was used for side view imaging. To adjust the size of laser spot areas, three long working distance objective lenses with magnifications of 20×, 50×, and 100× were applied to focus laser beam onto the sample surface. A frame rate of 100 kfps was used for imaging.

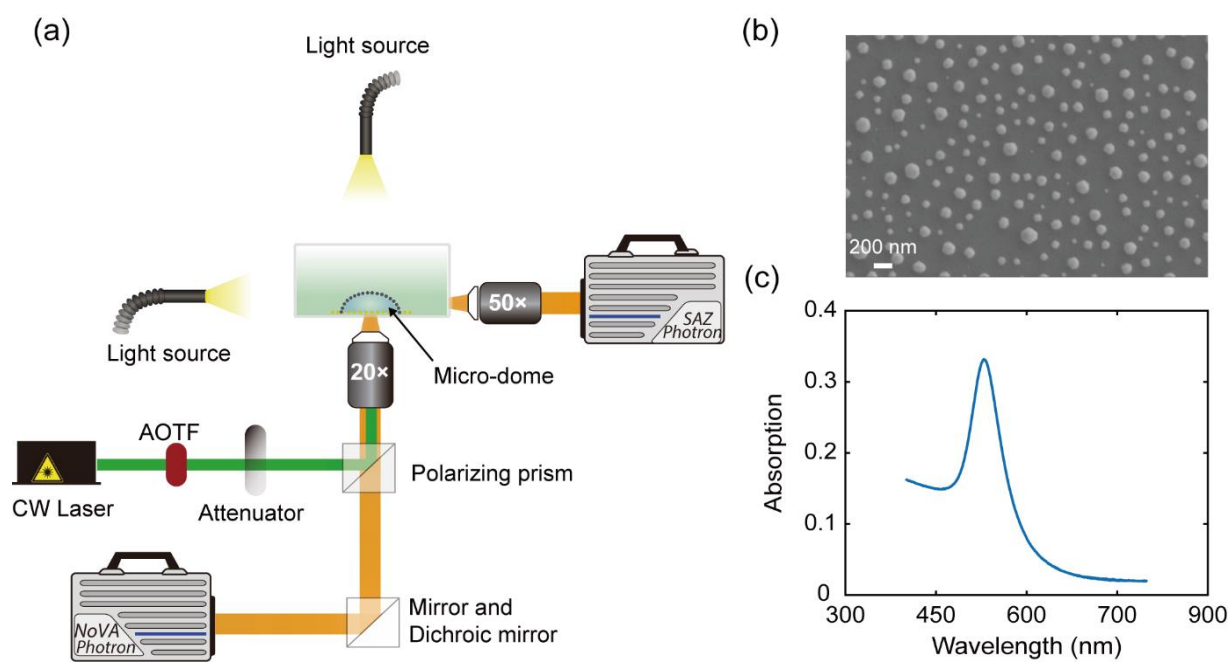

**Figure 7.** (a) Optical setup designed for experiments. (b) A scanning electron microscope (SEM) image of the GNP decorated sample surface. (c) The absorption spectrum of the GNP decorated sample surface.

*Preparation of the GNP decorated sample surfaces*

The GNP decorated sample surface was fabricated through a thermal dewetting method. To do so, a cleaned fused silica substrate was first coated with a 10 nm thick gold film using a sputter coater (108 Auto, Cressington, UK). After that, the sample surface was transferred to a muffle furnace for thermal dewetting under an annealing temperature of 1000 ºC for 1 hour. After that, the sample surface was removed from the muffle furnace and was cooled down to the room temperature. Figure 7b shows a scanning electron microscope (SEM) image of the fabricated sample surface. It shows that GNPs with an average diameter of 122 nm were obtained on the fused silica substrate. The optical absorption spectrum of the fabricated substrate was measured by using a UV-Vis spectrophotometer (U-3900H, Hitachi, Japan). The obtained absorption spectrum of the sample is shown in Figure 7c. It clearly shows that a peak absorption value of 0.33 was achieved at the wavelength of 529 nm, which is very close to that of 532 nm of the applied laser. For details of the sample preparation, please refer to one of our recent publications.[40]

*Numerical simulations of temperature and fluid fields around micro-domes*

The numerical simulations were conducted to construct temperature and fluid fields in the vicinity of micro-domes by using a commercial software (COMSOL Version 6.0, COMSOL Co., Ltd., Sweden). In the simulations, laser irradiation was modeled as a constant heat flux with Gaussian distribution. The heat flux is deposited on the substrate and acts as a heat source. For simplicity, the simulation region was modeled in a two-dimensional axisymmetric cylindrical coordinate system with a radius of 1.0 mm. The height for both water and fused silica in the simulation region is 1.0 mm. The thermodynamic parameters of water, fused silica, and iron oxide, such as heat capacity, density, thermal conductivity and so on, were from the material library of the software. The thermal conductivity of iron oxide was set as 12 W/(m·K) [41].

Simulations were carried out on three kinds of microstructures, plasmonic bubbles, hollow micro-shells, and solid micro-domes. The results of the first two kinds of simulations were included in the Supporting Information. They are used to investigate the temperature and fluid fields around a plasmonic bubble or a micro-shell. The simulation on solid micro-domes were used to investigate the spatial and temporal evolution of temperature fields around micro-domes. This helps to model the growth dynamics of micro-domes.

## Supporting Information

Supporting Information is available from the Wiley Online Library or from the author.

## Acknowledgements

Authors acknowledge the financial support from National Natural Science Foundation of China (grant No. 52075029) and "the Fundamental Research Funds for the Central Universities"

## Conflict of Interest

The authors declare no conflict of interest.

# References

[1]     J. R. Adleman, D.A. Boyd, D. G. Goodwin, D. Psaltis, *Nano Lett.*, **2009**, *9*, 4417.

[2]     A. Shakeri-Zadeh, H. Zareyi, R. Sheervalilou, S. Laurent, H. Ghaznavi, H. Samadian, *J. Control. Release*, **2021**, *330*, 49.

[3]     D. Lapotko, *Nanomedicine*, **2009**, *4*, 813.

[4]     H. T. Kim, H. Bae, Z. J Zhang, A. Kusimo, M. Yu, *Biomicrofluidics*, **2014**, *8*, 054126.

[5]     O. Neumann, A. S. Urban, J. Day, S. Lal, P. Nordlander, N. J. Halas, *ACS Nano*, **2013**, *7*, 42.

[6]     S. Fujii, R. Fukano, Y. Hayami, H. Ozawa, E. Muneyuki, N. Kitamura, M. A. Haga, *ACS Appl. Mater. Interfaces*, **2017**, *9*, 8413.

[7]     F. Tantussi, G. C. Messina, R. Capozza, M. Dipalo, L. Lovato, F. De Angelis, *ACS Nano*, **2018**, *12*, 4116.

[8]     L. Mohan, R. Hattori, H. P. Zhang, Y. Matsumura, T. S. Santra, T. Shibata, S. J. Ryu, M. Nagai, *Surfaces and Interfaces*, **2022**, *30*, 101820.

[9]     S. C. Yang, W. J. Fischer, T. L. Yang, *Appl. Phys. Lett.*, **2016**, *108*, 183105.

[10]    X. L. Li, Y. L. Wang, M. E. Zaytsev, G. Lajoinie, H. L. The, J. G. Bomer, J. C. T. Eijkel, H. J. W. Zandvliet, X. H. Zhang, D. Lohse, *J. Phys.Chem. C*, **2019**, *123*, 23586.

[11]    B. L. Zeng, Y. L. Wang, M. E. Zaytsev, Ch. L. Xia, H. J. W. Zandvliet, D. Lohse, *Phys. Rev. E*, **2020**, *102*, 063109.

[12]    M. E. Zaytsev, G. Lajoinie, Y. L. Wang, D. Lohse, X. H. Zhang, *J. Phys. Chem. C*, **2018**, *122*, 28375.

[13]    Y. L. Wang, M. E. Zaytsev, G. Lajoinie, T. H. The, J. C. T. Eijkel, A. Van Den Berg, M. Versluis, B. M. Weckhuysen, X. H. Zhang, H. J. W. Zandvliet, D. Lohse, *Proc. Natl. Acad. Sci.*, **2018**, *115*, 7676.

[14]    Y. L. Wang, M. E. Zaytsev, H. L. The, J. C. T. Eijkel, H .J. W. Zandvliet, X. H. Zhang, D. Lohse, *ACS Nano*, **2017**, *11*, 2045.

[15]    Y. Yamamoto, S. Tokonami, T. Iida, *ACS Appl. Bio Mater.*, **2019**, *2*, 1561.

[16]    B. L. Zeng, K. Lo. Chong, Y. L.Wang, C., Li Diddens, X. L. Li, M. Detert, H. J. W. Zandvliet, D. Lohse, *Proc. Natl. Acad. Sci.*, **2021**, *118*, e2103215118.

[17]   S. Jones, D. Andren, T J. Antosiewicz, A. Stilgoe, H. Rubinsztein-Dunlop, M. Kall, *ACS Nano*, **2020**, *14*, 17468.

[18]   K. Setoura, S. Ito, H. Miyasaka, *Nanoscale*, **2017**, *9*, 719.

[19]   L. H. Lin, X. L. Peng, Z. M. Mao, W. Li, M. N. Yogeesh, B. B. Rajeeva, E. P. Perillo, A. K. Dunn, D. Akinwande, Y. B. Zheng, *Nano Lett.*, **2016**, *16*, 701.

[20]   B. B. Rajeeva, L. H. Lin, E. P. Perillo, X. L Peng, W. W. Yu, A. K. Dunn, Y. B. Zheng, *ACS Appl. Mater. Interfaces*, **2017**, *9*, 16725.

[21]   B. B. Rajeeva, M. A. Alabandi, L. H. Lin, E. P. Perillo, A. K. Dunn, Y. B. Zheng, *J. Mater. Chem. C*, **2017**, *5*, 5693.

[22]   E. H. Hill, C. Goldmann, C. Hamon, M. Herber, *J. Phys. Chem. C*, **2022**, *126*, 7622.

[23]   S. Moon, Q. S. Zhang, D. Z. Huang, S. Senapati, H. C. Chang, E. Lee, T. F. Luo, *Adv. Mater. Interfaces*, **2020**, *7*, 200597.

[24]   E. Edri, N. Armon, E. Greenberg, E. Hadad, M. R. Bockstaller, H. Shpaisman, *ACS Appl. Mater. Interfaces*, **2020**, *12*, 22278.

[25]   T. Uwada, S. Fujii, T. Sugiyama, A. Usman, A. Miura, H. Masuhara, K. Kanaizuka, M. Haga, *ACS Appl. Mater. Interfaces*, **2012**, *4*, 1158.

[26]   B. Roy, M. Arya, P. Thomas, J. K. Jurgschat, K. V. Rao, A. Banerjee, C. M. Reddy, S. Roy, *Langmuir*, **2013**, *29*, 14733.

[27]   N. Armon, E. Greenberg, M. Layani, Y. S. Rosen, S. Magdassi, H. Shpaisman, *ACS Appl. Mater. Interfaces*, **2017**, *9*, 44214.

[28]   E. Edri, N. Armon, E. Greenberg, S. Moshe-Tsurel, D. Lubotzky, T. Salzillo, I. Perelshtein, M. Tkachev, O. Girshevitz, H. Shpaisman, *ACS Appl. Mater. Interfaces*, **2021**, *13*, 36416.

[29]   H. Kong, H. Kim, S. Hwang, J. Mun, J. Yeo, *ACS Appl. Nano Mater.*, **2022**, *5*, 4102.

[30]   B. B. Rajeeva, P. Kunal, P. S. Kollipara, P. V. Acharya, M. Joe, M. S. Ide, K. Jarvis, Y. Y. Liu, V. Bahadur, S. M. Humphrey, Y. B. Zhang, *Matter*, **2019**, *1*, 1606.

[31]   H. M. L. Robert, F. Kundrat, E. Bermudez-Urena, H. Rigneault, S. Monneret, R. Quidant, J. Polleux, G. Baffou, *ACS Omega*, **2016**, *1*, 2.

[32]   B. B. Rajeeva, Z. L. Wu, A. Briggs, P. V. Acharya, S. B. Walker, X. L. Peng, V. Bahadur, S. R. Bank, Y. B. Zheng, *Adv. Optical Mater.*, **2018**, *6*, 1701213.

[33]  S. Ghosh, S. Das, S. Paul, P. Thomas, B. Roy, P. Mitra, S. Roy, A. Banerjee, *J. Mater. Chem. C*, **2017**, *5*, 6718.

[34]  E. Greenberg, N. Armon, O. Kapon, M. Ben-Ishai, H. Shpaisman, *Adv. Mater. Interfaces*, **2019**, *6*, 1900541.

[35]  N. Armon, E. Greenberg, E. Edri, A. Kenigsberg, S. Piperno, O Kapon, O Fleker, I Perelshtein, G Cohen-Taguri, I Hod, H. Shpaisman, *Chem. Commun.*, **2019**, *55*, 12773.

[36]  K. Wieczorek-Ciurowa, A. J. Kozak, *J. Therm. Anal. Cal.*, **1999**, *58*, 647.

[37]  C. W. Lee, P. C. Wu, I. L. Hsu, T. M. Liu, W. H. Chong, C. H. Wu, T. Y. Hsieh, L. Z. Guo, Y. Tsao, P. T. Wu, J. S. Yu, P. J. Tsai, H. S. Huang, Y. C. Chuang, C. C. Huang, *Small*, **2019**, *15*, 1805086.

[38]  Q. L. Kang, Y. Z. Qin, J. W. Shi, B. R. Xiong, W. Y. Tang, F. Gao, Q. Y. Lu, *J. Colloid Interface Sci.*, **2022**, *622*, 780.

[39]  I. W. Un, Y. Sivan, *Nanoscale*, **2020**, *12*, 17821.

[40]  CH. L. Xia, R. Wang, P. W. Zhu, F. L. Wang, L. H. Dong, H. M. Wang, Y. L. Wang, *Surfaces and Interfaces*, **2023**, *36*, 102591.

[41]  M. Takeda, T. Onishi, S. Nakakubo, S. Fujimoto, *Materials Transactions*, **2009**, *50*, 2242.

# Supporting Information

## Direct Laser Writing of Surface Micro-Domes by Plasmonic Bubbles

Lihua Dong[1,4], Fulong Wang[1], Buyun Chen[1], Chenliang Xia[1,4], Pengwei Zhu[2], Zhi Tong[2], Huimin Wang[2], LijunYang[3,4], and YuliangWang[1,4,*]

[1] School of Mechanical Engineering and Automation, Beihang University, 37 Xueyuan Rd., Haidian District, Beijing 100191, China

[2] National Center for Materials Service Safety, University of Science and Technology Beijing, Beijing 100083, China

[3] School of Astronautics, Beihang University, Beijing 100191, China

[4] Ningbo Institute of Technology, Beihang University, Ningbo 315832, China

* E-mail: wangyuliang@buaa.edu.cn

# 1. The formation process of a micro-dome

**Figure S1** shows the detailed formation process of a micro-dome. In the first 30 s of the laser irradiation, a wiggling bubble was first nucleated ($t = 6$ s) and detached from the substrate ($t = 10.8$ s). The process may repeat for several times until a wiggling bubble gradually stabilized on the substrate ($t = 33$ s). With time, the gradually stabilized bubble became opaque ($t = 55$ s) and a micro-dome was formed. When the laser was switched off at $t = 600$ s, a micro-dome was formed with scattered nanoparticles.

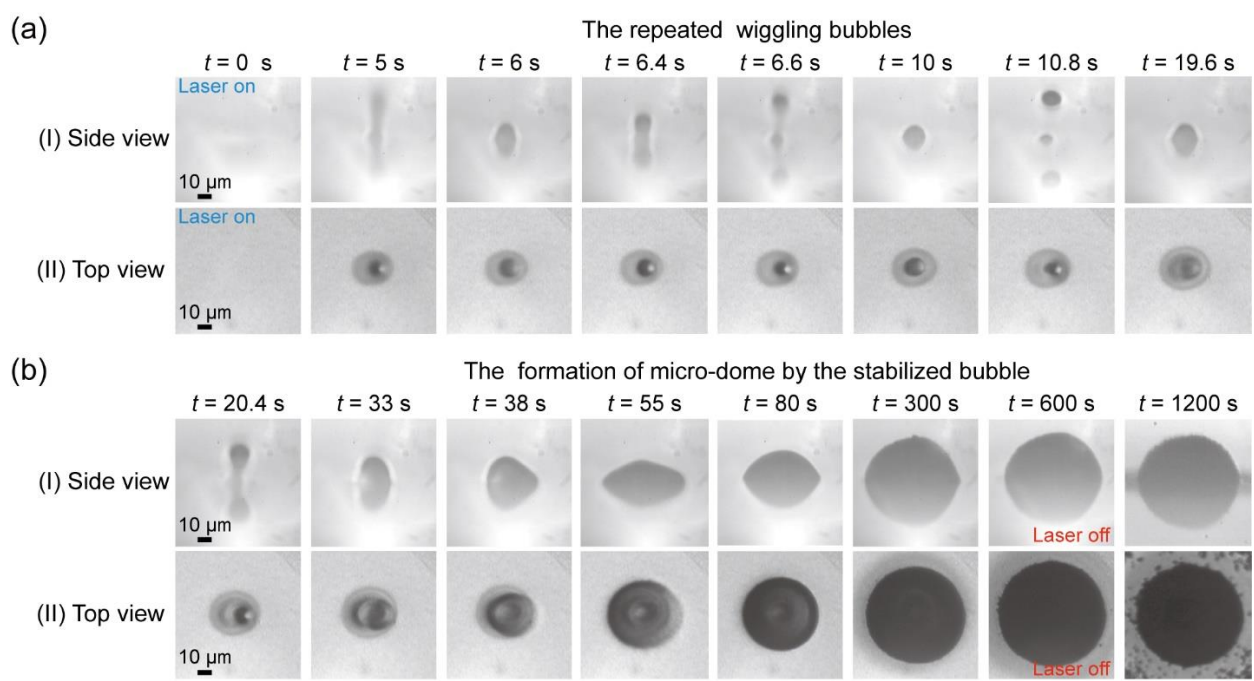

**Figure S1.** Synchronized side view and top view images of repeatedly nucleated wiggling bubbles ($t = 0 - 33$ s) and a micro-dome ($t = 38 - 600$ s) under laser irradiation on a GNP decorated sample surface ($P_l = 27.85$ mW, $c = 1.0\%$, and $m = 20\times$) The first 30 s of laser irradiation is the phase of the repeatedly nucleated wiggling bubbles. After that, a wiggling bubble gradually stabilized on the substrate. By taking this bubble as a template, a micro-dome structure gradually formed.

# 2. Mechanism of the micro-dome formation

**Figure S2** shows the optical and SEM images for several micro structures obtained with different laser irradiation period $t_d$ of 50, 150, 180, 200, 400, and 600 s. Figure S2(a) and S2(b) are side view and top view images of these microstructures, while Figure S2(c) and S2(d) are isometric view and top view images of the same micro-structures. From the images, one can see that at beginning, it first forms a micro-rim ($t_d = 50$ s), followed by a micro-shell structure ($t_d = 150$, 180, and 200 s). With time, the micro-shell is completely enclosed, forming a micro-dome ($t_d = 400$ s). The diameter of the obtained micro-domes increases with time ($t_d = 600$ s).

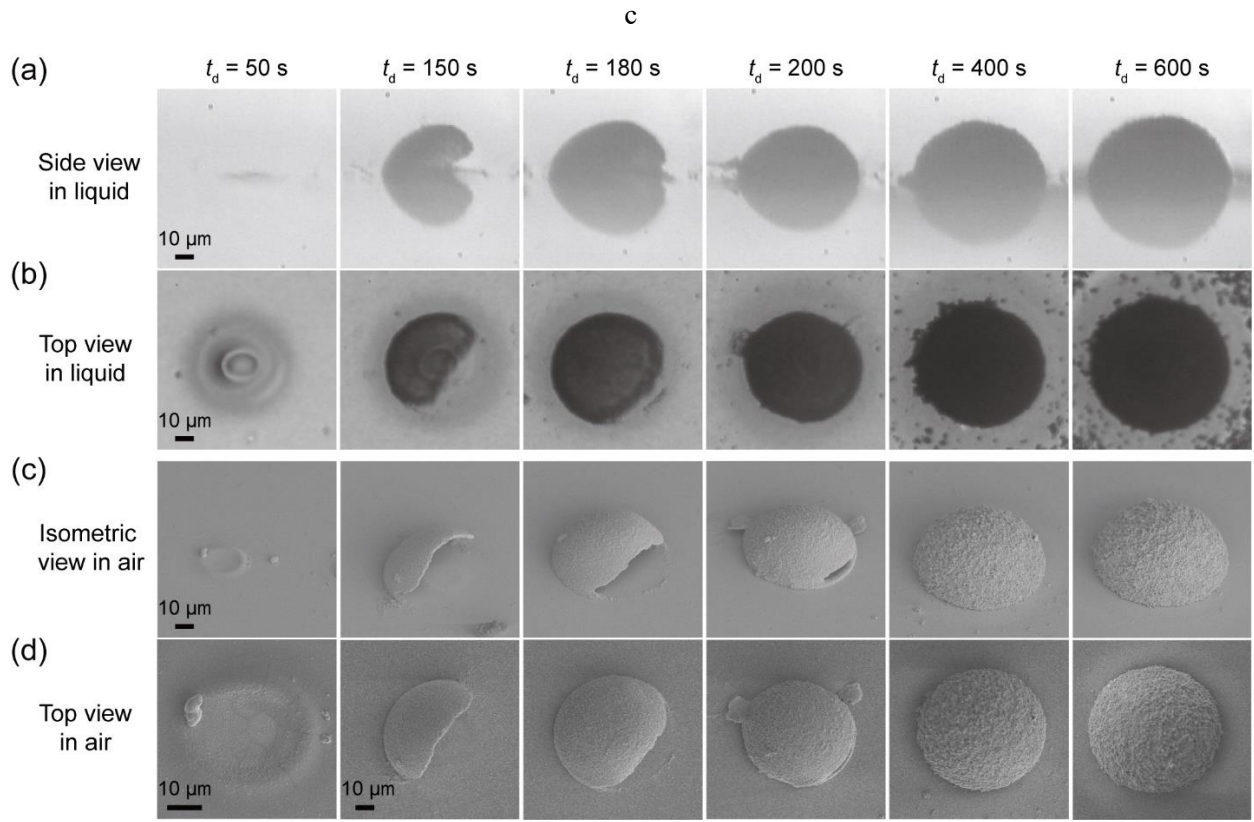

**Figure S2.** Optical and SEM images revealing the step by step formation of micro-domes ($P_l = 27.85$ mW, $c = 1.0\%$, and $m = 20\times$). Side view (a) and top view (b) optical images taken in liquid after laser irradiation of different periods of time $t_d$. Isometric (c) and top view (d) SEM images of the same microstructures shown in (a) and (d).

**Figure S3** shows the evolution of micro-rims formed at the beginning of laser irradiation. Figure S3(a, I) and Figure S3(a, II) show the SEM images and corresponding top view optical images of several micro-rims obtained with different laser irradiation period $t_d$. Figure S3(b) depicts the footprint diameter $L_r$ of the micro-rims as a function of $t_d$. It clearly shows that $L_r$ increases with $t_d$, with its maximum value equals to laser diameter. This indicates that the micro-rims are laser spot confined microstructures.

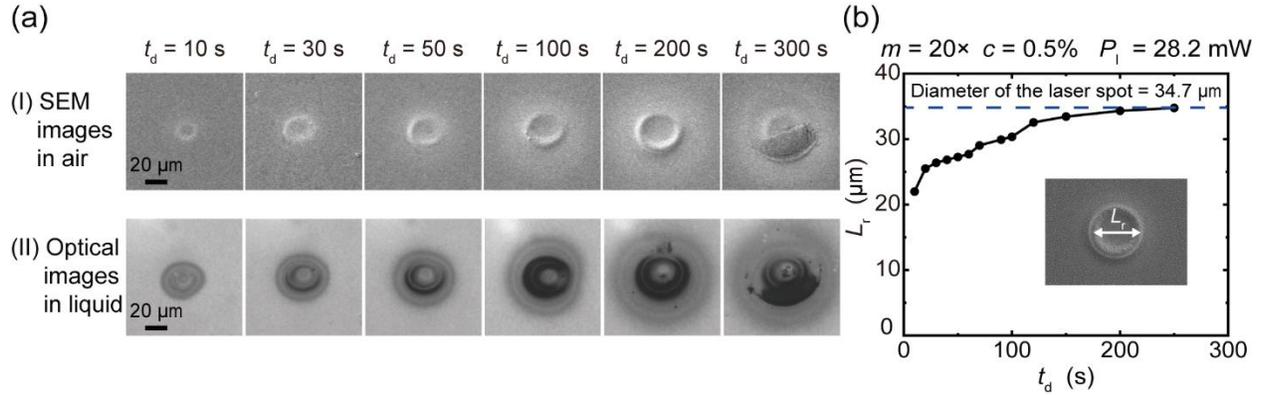

**Figure S3.** Dependence of micro-rim size on laser irradiation period $t_d$ ($P_l$ = 28.2 mW, $c$ = 0.5%, and $m$ = 20×). (a) SEM images in air (I) and corresponding optical images in liquid (II) of several micro-rims generated with different laser irradiation period $t_d$. (b) The footprint diameter $L_r$ of micro-rims as a function of $t_d$. The size of micro-rims is confined by the laser spot.

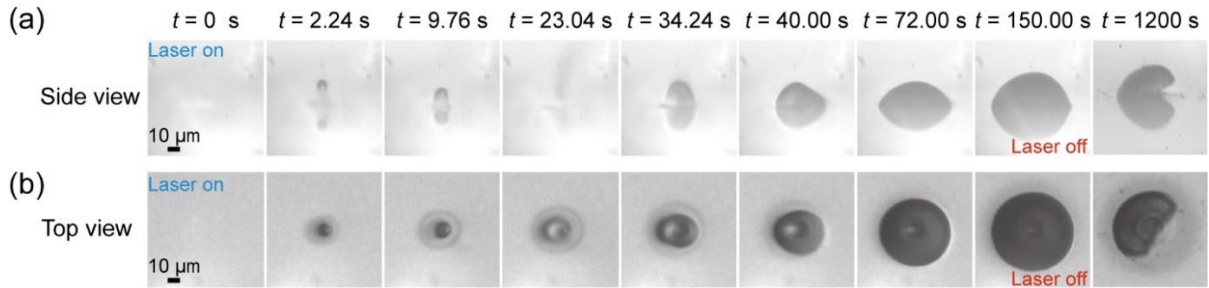

**Figure S4.** Sequentially captured and synchronized side view (a) and top view (b) optical images of a laser spot area, indicating the generation process of a semi-closed micro-shell structure.

**Figure S4** shows the sequentially captured and synchronized side view and top view optical images of a semi-closed microstructure at $P_l$ = 27.85 mW and $c$ = 1.0% with a 20× focusing objective lens. At about $t$ = 34.24 s, a plasmonic bubble was pinned at the substrate. Initially, it kept wiggling ($t$ = 40.0 s). With time, it was stabilized. The ferric oxide particles were deposited on surface of the stabilized bubble, leading to the increased size of the obtained microstructure ($t$ = 72.0 s). When the laser was switched off at 150 s, some of the particles were slipped off from the microstructure, due to the weak bonding between the particles and the surface of the microstructure.

## 3. Simulation of the temperature fields around plasmonic microbubbles and micro-shells

To investigate the surface temperature and the fluid flows around microbubbles and micro-shells, the numerical simulations were carried out by the COMSOL software. In the simulations, the radii of the microbubble and the micro-shell were set as 22 $\mu$m and 30 $\mu$m, respectively, according to that obtained experimentally. The thickness of the micro-shell was set 8.0 $\mu$m. The simulation results are depicted in **Figure S5**. For the microbubble, it shows that the temperature at the three-phase contact line can reach 387 K, which is much higher than the required temperature for thermal decomposition of ferric nitrate solution (Figure S5(a)). Moreover, it also exists a strong upward convective flow (up to 500 mm/s, Figure S5(b)). This facilitates the upward migration of ferric oxide nanoparticles from the three-phase contact line along bubble surface.

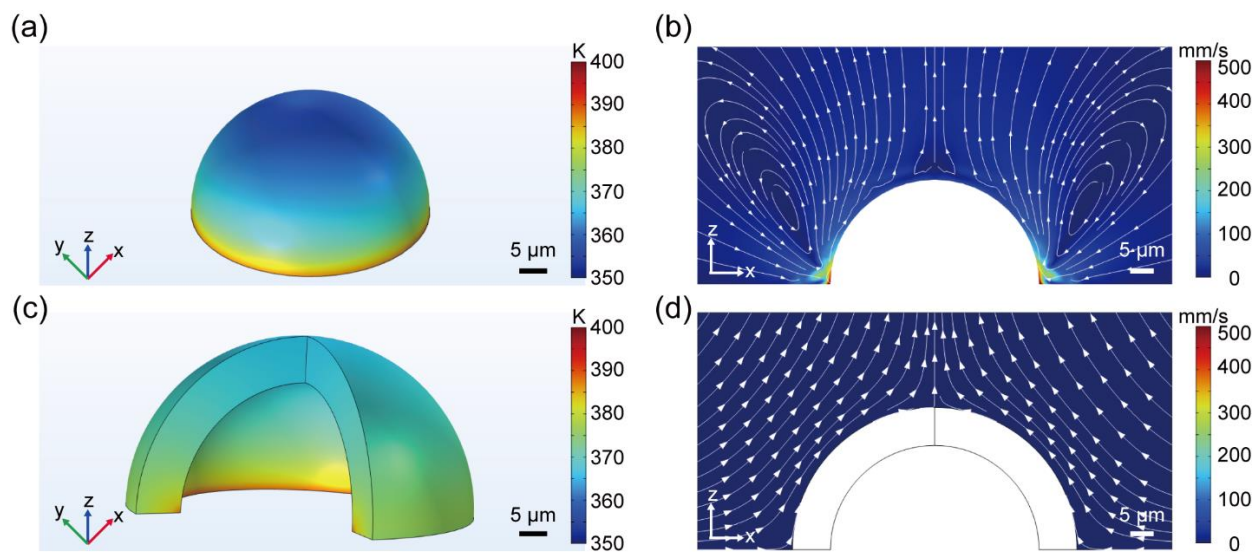

**Figure S5.** COMSOL simulation of temperature and fluid fields around a microbubble and a micro-shell. (a) and (b) are obtained temperature and fluid fields around a microbubble surface, respectively. (c) and (d) are temperature and fluid fields around a micro-shell.

For the micro-shell structure, the temperature at its apex is about 360 K, which is also high enough to trigger thermal decomposition of ferric nitrate solution (Figure S5(c)). This guarantees the continuous nanoparticle formation on the micro-shell surface. We also find that the flow rate around the micro-shell is significantly decreased compared to that of the microbubble (Figure S5(d)). This indicates that the growth of micro-shells is believed to be dominated by direct nanoparticle nucleation on the micro-shell surface, not the migration from the three-phase contact line.

## 4. Reproducibility of micro-dome growth dynamics

**Figure S6(a)** and Figure S6(b) show the radii $R_d$ and the volume $V_d$ of the repeatedly fabricated micro-domes as a function of time $t$ with $P_l = 7.85$ mW and $m = 50\times$ at different ferric nitrate concentrations. It exhibits a good reproducibility, regardless of the ferric nitrate concentration. Moreover, Figure S6(c) is the enlarged plot for the selected area in Figure S6(b). It clearly shows that $V_d$ linearly increases with $t$, exhibiting a reliable $V_d(t) \propto t$ or $R_d(t) \propto t^{1/3}$ dependence.

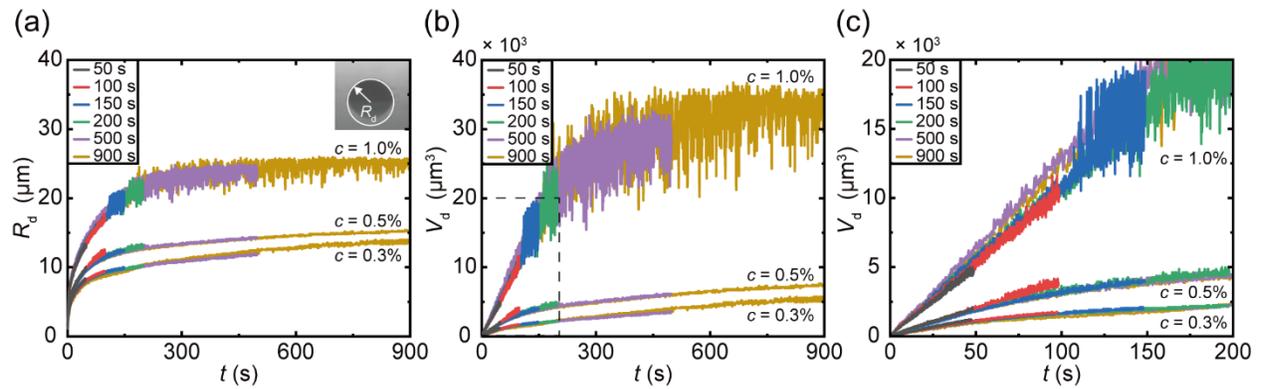

**Figure S6.** Growth dynamics of micro-domes with different ferric nitrate concentration and laser irradiation periods at $P_l = 7.85$ mW and $m = 50\times$. (a) Radii of curvature $R_d$ and (b) volume of $V_d$ of micro-domes as functions of $t$ along laser irradiation. It exhibits a good reproducibility. (c) Enlarged plot for the selected area in (b). It clearly shows that $V_d$ linearly increases with $t$ for $t \leq 200$ s.

## 5. Supplementary movies

The supplementary materials include four movies:

Movie S1 shows the growth dynamics of the jet plumes with different laser irradiation periods $t_d$ at $P_l = 48.47$ mW and $c = 10.0\%$ by using a 20× focusing objective lens.

Movie S2 shows the dynamics of a plume from bottom view at $P_l$ = 55.0 mW and $c$ = 10.0% with a 20× focusing objective lens, a shell is first ejected and then fall off on the substrate.

Movie S3 is the synchronized video from side view and bottom view, showing the formation process of a micro-dome at $P_l$ = 27.85 mW and $c$ = 1.0% with a 20× focusing objective lens.

Movie S4 is the synchronized video from side view and bottom view, showing the formation process of a semi-closed micro-shell from the side and bottom views at $P_l$ = 27.85 mW and $c$ = 1.0% with a 20× focusing objective lens.